\documentclass[11pt]{article}
\pdfoutput=1 
\usepackage{jheppub} 
\usepackage{bbm,bm,graphicx,mathtools,color,hyperref,slashed}
\usepackage{amssymb,amsmath}
\usepackage{youngtab}
\usepackage[T1]{fontenc}

\graphicspath{{./Figures/}}

\newcommand{\diff}{\mathrm{d}}
\newcommand{\p}{\partial}

\newcommand{\Diff}{{\mathcal{D}}}

\newcommand{\be}{\begin{equation}}      
\newcommand{\ee}{\end{equation}}      
\newcommand{\bea}{\begin{eqnarray}}      
\newcommand{\eea}{\end{eqnarray}}

\newcommand{\tr}{\mathrm{tr}}

\newcommand{\im}{\mathrm{i}}

\newcommand{\Gr}{\mathrm{Gr}}

\newcommand{\calH}{\mathcal{H}}

\newcommand{\calZ}{\mathcal{Z}}

\newcommand{\R}{\mathbb{R}}

\title{Quantum Distillation of Hilbert Spaces, Semi-classics and  Anomaly Matching}
\author[a,b]{Gerald V. Dunne,}
\emailAdd{gerald.dunne@uconn.edu}
\affiliation[a]{Kavli Institute for Theoretical Physics, University of California Santa Barbara, CA 93106 USA}
\affiliation[b]{Department of Physics, University of Connecticut, Storrs CT 06269, USA}
\author[a,c]{Yuya Tanizaki,}
\affiliation[c]{RIKEN BNL Research Center, Brookhaven National Laboratory, Upton, NY 11973 USA}
\emailAdd{yuya.tanizaki@riken.jp}
\author[a,d]{Mithat \"{U}nsal}
\affiliation[d]{Department of Physics, North Carolina State University, Raleigh, NC 27695, USA}
\emailAdd{unsal.mithat@gmail.com}

\abstract{
A symmetry-twisted boundary condition of the path integral provides a suitable framework for the semi-classical analysis of nonperturbative quantum field theories (QFTs), and we reinterpret it from the viewpoint of the Hilbert space. 
An appropriate twist with the unbroken symmetry can potentially produce huge cancellations among excited states in the state-sum, without affecting the ground states; we call this effect ``quantum distillation''. 
Quantum distillation can provide the underlying mechanism for adiabatic continuity, 
by preventing a phase transition under $S^1$ compactification. 
We revisit this point via the 't Hooft anomaly matching condition when it constrains the vacuum structure of the theory on $\mathbb{R}^d$ and upon compactification. 
We show that  there is a precise relation between the  persistence of the anomaly upon compactification, the Hilbert space quantum distillation, and the semi-classical analysis of the corresponding symmetry-twisted path integrals.
We motivate quantum distillation in quantum mechanical examples, and then study its non-trivial action in QFT, with the example of the 2D Grassmannian sigma model $\mathrm{Gr}(N,M)$. 
We also discuss the connection of quantum distillation with large-$N$ volume independence and flavor-momentum transmutation.
\vskip 3cm
\noindent
NSF-ITP-18-006\\
RBRC-1271
}

\begin{document}
\maketitle

\section{Introduction and sketch of the idea}\label{sec:introduction}
In this paper, we connect the semi-classical analysis of asymptotically free quantum field theory in the calculable weak-coupling domain and with adiabatic continuity~\cite{Dunne:2016nmc, Dunne:2012ae,Unsal:2007jx,Unsal:2008ch} (see also 
~\cite{ Unsal:2007vu, Kovtun:2007py,  Shifman:2008ja, Shifman:2009tp, Cossu:2009sq, Cossu:2013ora, Argyres:2012ka, Argyres:2012vv, Dunne:2012zk, Poppitz:2012sw, Anber:2013doa, Basar:2013sza, Cherman:2013yfa, Cherman:2014ofa, Misumi:2014raa, Misumi:2014jua, Misumi:2014bsa, Dunne:2015ywa,Misumi:2016fno, Cherman:2016hcd, Fujimori:2016ljw, Sulejmanpasic:2016llc, Yamazaki:2017ulc, Buividovich:2017jea, Aitken:2017ayq})  with the 't Hooft anomaly matching between the short distance (UV) and long distance (IR) effective theories~\cite{tHooft:1979rat, Gaiotto:2017yup}  (see also \cite{Frishman:1980dq, Coleman:1982yg, Vishwanath:2012tq, Wen:2013oza, Kapustin:2014lwa, Kapustin:2014zva, Cho:2014jfa, Wang:2014pma, Witten:2015aba,Seiberg:2016rsg,Witten:2016cio, Tachikawa:2016cha, Tachikawa:2016nmo, Gaiotto:2017yup,Wang:2017txt, Tanizaki:2017bam, Komargodski:2017dmc, Komargodski:2017smk, Cho:2017fgz,Shimizu:2017asf, Wang:2017loc, Metlitski:2017fmd,Kikuchi:2017pcp, Gaiotto:2017tne, Tanizaki:2017qhf, Tanizaki:2017mtm, Guo:2017xex}). 
We motivate our approach using ideas from graded representation theory of Hilbert space.
In 2d QFT, it has been shown that adiabatic circle compactification with special spatial twisted boundary conditions permits well-controlled semi-classical analysis of features such as symmetry breaking, revealing the correct theta-angle dependence (when it exists), mass gap, Borel plane and UV-IR renormalon structure. This approach applies to QFTs with instantons and theta angles, such as $O(3)$, $\mathbb {CP}^{N-1}$, and $\Gr(N, M)$  sigma models~\cite{Dunne:2012ae, Dunne:2012zk,Misumi:2014jua, Misumi:2014bsa, Dunne:2015ywa}, as well as to theories without instantons or a  $\Theta$-term, such as $O(N)$ with $N>3$, and the Principal Chiral Model~\cite{Cherman:2013yfa, Dunne:2015ywa, Buividovich:2017jea}. 
In both  classes of QFTs, semi-classical analysis reveals the existence of new saddles \cite{Dunne:2016nmc}. Our new Hilbert space approach gives extra physical insight into the special twisted boundary conditions and their relation to the vacuum structure of the theory in decompactification limit.
In the first class of theories, those with instantons and $\Theta$-angles, these unique boundary conditions are the ones for which a mixed anomaly polynomial, despite being associated with a zero form symmetry, persists upon compactification. 

In general, Hilbert spaces of QFTs and quantum mechanical (QM) systems are  ``big places''.  For example, even for a modest QM spin system with  50 spin-$\frac{1}{2}$ particles,
the dimension of the Hilbert space is very large: $\mathcal{H}$, ${\rm dim}( \mathcal{H}) = 2^{50}\approx 10^{15}$. For the computation of certain properties of the system, not every state is as important as the others, as is familiar for example from the success of variational approaches to low energy and long-distance properties.
In this work, motivated by ideas from graded representation theory,  we analyze the structure of the Hilbert space, and try to extract the most important ground state contributions out of the vast Hilbert space. 
We show that this is equivalent to using the symmetry-twisted partition function instead of the thermal one. 
The idea is  reminiscent of the supersymmetric (Witten) index~\cite{Witten:1982df, Witten:1982im,Witten:1986bf}, but it works equally well for non-supersymmetric theories,  including purely bosonic theories. In supersymmetric systems, a natural grading is the $\mathbb{Z}_2$ fermionic number, $(-1)^F$. This grading  distinguishes bosonic states from fermionic states, leading to large cancellations between degenerate states under supersymmetry, and turns a thermal state sum into a graded state-sum  $\calZ(L)= \mathrm{tr}_{\mathcal{H}}\left[\exp(-L  \widehat{H}) (-1)^F \right]$, which has no dependence on $L$.  We are also motivated by the success of twisted partition functions in probing the low energy and non-perturbative properties of supersymmetric quantum field theories  \cite{Nekrasov:2002qd,Nekrasov:2003rj,Nekrasov:2009rc}.
In a purely bosonic theory, or in general non-supersymmetric theories, we seek a similar operation that distills the ground state from excited states: we call this procedure \textit{quantum distillation}.

As a general setup, let us consider a QFT with the global symmetry $G$, and denote its Hilbert space and Hamiltonian as $\mathcal{H}$ and $\widehat{H}$, respectively. 
If the symmetry $G$ is spontaneously broken to $G_0$, the Hilbert space is decomposed into superselection sectors labeled by the coset space $G/G_0$:
\be
\calH=\bigoplus_{v\in G/G_0}\calH^{(v)}, 
\ee
where the matrix elements of any local operators between $\calH^{(v)}$ and $\calH^{(v')}$ are zero in the infinite volume limit ($v\not=v'$). Since $\calH^{(v)}$ and $\calH^{(v')}$ are related by the broken symmetry $G$, they have the same energy spectrum. 
Assuming the absence of accidental degeneracy, energy eigenstates of each $\calH^{(v)}$ with energy $E$ are in some irreducible representation $\mathcal{R}_E$ of the unbroken symmetry $G_0$, and especially the ground state is singlet which we denote as  $\mathcal{R}_{0}$:  
\begin{align}
\mathcal{H}^{(v)}
 \simeq \mathcal{R}_{0}  \oplus \bigoplus_{E>0} {\mathcal R}_E. 
\end{align}
The ordinary partition function treats states within a given representation ${\mathcal R}_E$ exactly in the same way, since each state appears once with its Boltzmann weight:
\begin{eqnarray}
\calZ(\beta)=\mathrm{tr}_{\mathcal{H}}\left[\exp(-\beta \widehat{H})\right] 
=\mathrm{vol}(G/G_0) \left(1+\sum_{E>0} \mathrm{dim}({\mathcal{R}_E}) \left[\exp(-\beta\, E)\right]\right). 
\label{eq:thermal}
\end{eqnarray}
For a system with large global symmetries, 
${\rm dim}( {\mathcal R}_E) $ can grow very quickly as a function of $E$. Therefore, in order to extract  ground state properties, such as the ground state degeneracy $\mathrm{vol}(G/G_0)$, we have to take the limit $\beta\to\infty$, which is typically the strongly-coupled regime of interesting field theories. 

Let $\widehat{Q}_i$ denote the generators of the maximal torus of $G_0$, so that $[\widehat{Q}_i, \widehat{Q}_j] = [\widehat{Q}_i, \widehat{H}]=0$, and we introduce the symmetry-twist operator,
\be
\widehat{\Omega}=\exp\left(\im \sum_{j}\alpha_j \widehat{Q}_j\right). 
\ee
We define the symmetry-twisted partition function as:
\begin{align}
\calZ_\Omega(L)= \mathrm{tr}_{\mathcal{H}}\left[\exp(-L \widehat{H}) \widehat{\Omega} \right] 
=\mathrm{vol}(G/G_0)\left(1+\sum_{E>0}\mathrm{tr}_{\mathcal{R}_E}\bigl(\widehat{\Omega}\bigr)\exp(-LE)\right). 
\label{master}
\end{align}
Since  $[\widehat{Q}_i, \widehat{H}]=0 $, eigenstates of $\widehat{H}$ are also eigenstates of $\widehat{Q}_i$, and \eqref{master} provides pure phase assignments 
to states in each ${\mathcal R}_E$. These states in ${\mathcal R}_E $ are,  by symmetries,  degenerate, so the phases  provide an opportunity for the states within ${\mathcal R}_E  $ to 
interfere destructively, leading to  state-cancellations, so that $\mathrm{tr}_{\mathcal{R}_E}(\widehat{\Omega})\ll \dim(\mathcal{R}_E)$.  
When this is true, we can extract the ground-state properties from $\calZ_{\Omega}(L)$ without taking $L\to \infty$, providing  more useful information concerning the nonperturbative aspects of asymptotically free QFTs.\footnote{A real chemical potential does not cause a sign problem in the Hamiltonian formulation, but does  create a sign problem in the Euclidean path integral formulation. The grading in (\ref{master}) corresponds to imaginary chemical potentials.  In this case, it does not  create a  sign problem in the Euclidean path integral formulation, but it does cause a sign ``problem'' in the Hamiltonian formulation (the graded state-sum). However,  this Hamiltonian sign ``problem'' is actually useful.
Thus, the physical intuition behind real and imaginary chemical potentials are somehow opposite. The former creates preference of some states over the others by altering the magnitude of their Boltzmann weight, the latter does not change the magnitude of the Boltzmann weight, 
but attaches   a pure phase to the state according to its charge.}

\vspace{0.5cm}
\noindent
{\bf What is this good for?}    There are three related answers to this question. We also summarize these viewpoints in Fig.~\ref{fig:overview}: 

\begin{figure}[t]
\centering
\includegraphics[scale=0.7]{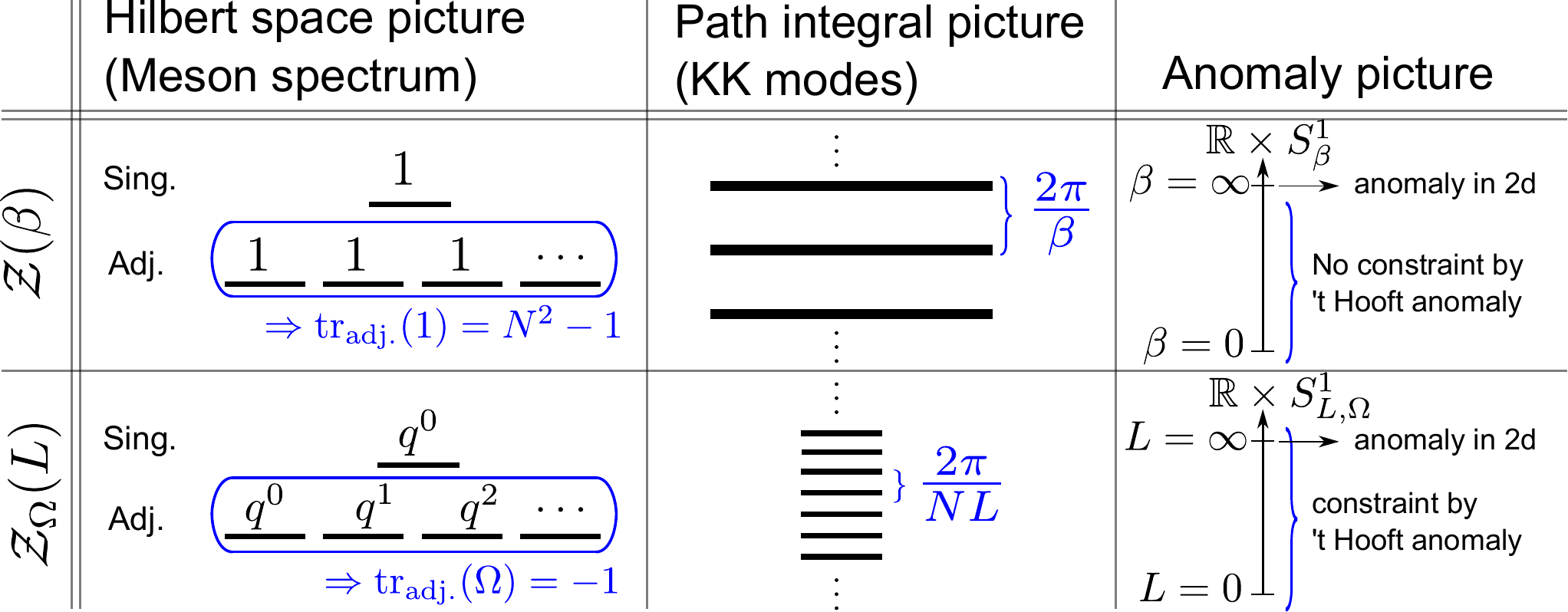}
\caption{Schematic illustration of the advantages of the symmetry-twisted partition function $\mathcal{Z}_{\Omega}(L)$ in comparison with the thermal partition function $\mathcal{Z}(\beta)$. In the left column, we count the degrees of freedom of mesons. Mesons give a large, $O(N^2)$ contribution to $\mathcal{Z}(\beta)$ since they are in the adjoint representation of the $SU(N)/\mathbb{Z}_N$ symmetry, while $\mathcal{Z}_{\Omega}(L)$ is affected only by $O(1)$. In the middle column, it is explained by the Kaluza-Klein (KK) or Matsubara decomposition of the fields, and the KK modes in $\mathcal{Z}_{\Omega}(L)$ are much denser than those of $\mathcal{Z}(\beta)$, due to the flavor-momentum transmutation. In the right column, we explain its consequence for the 't~Hooft anomaly of the theory, and the 't~Hooft anomaly in $2$ dimensions persists in $\mathcal{Z}_{\Omega}(L)$ for any $L$, while it survives in $\mathcal{Z}(\beta)$ only for $\beta\to\infty$. }
\label{fig:overview}
\end{figure} 

\begin{itemize}

\item {\bf Analyticity,  adiabatic continuity, and semi-classics:}
To study the ground state properties in a non-trivial asymptotically free QFT, a natural approach is to take advantage of the
weak coupling at small-$\beta$ (high temperature). However, at small-$\beta$,  $\calZ(\beta)$ is an extremely contaminated quantity. It has contributions from a huge tower of states with similar order of magnitude Boltzmann weights $e^{-\beta E}$. Furthermore,  in the presence of phase transitions,  
 $\calZ(\beta)$ is non-analytic as a function of $\beta$.
On the other hand, a  suitable $\calZ_\Omega(L)$ may be a relatively uncontaminated quantity, dominated by a few or even a single state,\footnote{In this sense, our approach has a similarity to tensor network  approach to QFT 
\cite{doi:10.1080/14789940801912366}.
Both may pick out a subset of important states in an otherwise  exponentially large Hilbert space.} 
due to large-cancellations in the state-sum \cite{Dunne:2012ae,Sulejmanpasic:2016llc} . As a result, it  may be analytic as a function of  $L$  and can continuously connect small and  large $L$ limits.
If this is the case,    the corresponding path integral   does not change dramatically as the radius is changed. At small-$L$ where the theory becomes weakly coupled, the corresponding path integral 
 encodes semi-classical non-perturbative information about the vacuum structure of the theory  which is 
  adiabatically connected to the strong coupling regime at large-$L$ \cite{Dunne:2016nmc}.
%
 Note that this  construction does not change the theory, vacuum or Hilbert space. We are simply probing its low energy properties with a different operator, leading to a generalization of  partition function, in which there may be potentially large cancellations between states,  and which may  be 
 analytic.

\item {\bf Quantum distillation and mixed anomalies:} Assume the quantum theory has a mixed anomaly between two symmetries $G_1$ and $G_2$. Further, assume that these are zero-form symmetries.
 It has recently been shown~\cite{Tanizaki:2017qhf}, and here we generalize this argument to the Grassmannian $\Gr(N,M)$ sigma model, that an anomaly polynomial that exists on $\R^d$ persists on  $\R^{d-1} \times S^1$ if and only if one uses a unique background holonomy associated with $G_1$. This is equivalent to putting a unique twisted boundary condition associated with the path integral formulation of the same theory, and correspondingly a unique (ideal) distillation of the Hilbert space.

\item{\bf Flavor-momentum transmutation in the path integral:} When a theory is compactified on a circle of radius $L$, and one uses the special 
 graded partition function $\calZ_\Omega(L)$,  the $G$ singlet observables in the theory exhibit volume independence at large rank$(G)$. 
 The Kaluza-Klein decomposition of modes in path integral involves the quantization of momentum  not in units of $\frac{2 \pi}{L}$, but rather in units of $\frac{2 \pi}{L \,{\rm rank}(G)}$, for large rank.
In other words, the flavor index transmutes to a momentum index. This provides the perturbative intuition behind the idea of large-$N$ volume independence \cite{Unsal:2010qh, Cherman:2016vpt}.
 
\end{itemize}

All these features of the symmetry-twisted partition function $\calZ_{\Omega}(L)$ provide special advantages for the study of the ground-state properties of asymptotically-free QFTs.

\section{Quantum distillation and graded partition functions in QM}\label{sec:intuition}
\label{sec:qm}

In this  section, we use two simple QM examples to explain the underlying physical intuition of the  graded (or symmetry-twisted) partition function. We first discuss the $N$-dimensional simple harmonic oscillator. The simplicity of this example should not deceive the reader. Since it is the global symmetry that matters for the discussion, this example illustrates some interesting effects relevant for non-trivial QFTs. Next, we discuss QM of a particle on the 
$\mathbb C \mathbb P^{N-1}$ manifold.

\subsection{$N$-dimensional isotropic simple harmonic oscillator}

Consider the $N$-dimensional isotropic simple harmonic oscillator
\be
\widehat{H}=\sum_{j=1}^{N}{1\over 2}\left(\widehat{p}_j^2+\widehat{x}_j^2\right)=\sum_{j=1}^{N}\left(\widehat{a}^{\dagger}_j \widehat{a}_j +\frac{1}{2}\right) \quad , 
\ee
which has a global $U(N)$ symmetry.
The canonical commutation relation is $[\widehat{x}_i,\widehat{p}_j]=\im \delta_{ij}$, and creation and annihilation operators are introduced as $\widehat{a}_i={1\over \sqrt{2}}(\widehat{x}_i+\im \widehat{p}_i)$ and $\widehat{a}^\dagger_i={1\over \sqrt{2}}(\widehat{x}_i-\im \widehat{p}_i)$, respectively. 
The ground state $|0\rangle$ is characterized by $\widehat{a}_i|0\rangle=0$ for all $i=1,\ldots,N$, and the Hilbert space is spanned by the Fock basis, 
\be
|\{n_i\}\rangle={1\over \sqrt{n_1! \cdots n_N !}}(\widehat{a}^{\dagger}_1)^{n_1}\cdots (\widehat{a}^{\dagger}_N)^{n_N}|0\rangle. 
\ee
The energy of the state $|\{n_i\}\rangle$ is 
\be
E_{\{n_i\}}=\frac{N}{2}+\sum_{i=1}^{N}n_i. 
\ee
The global $U(N)$ symmetry acts as: $\widehat{a}_j\mapsto U_{ij}\widehat{a}_j$, with $U\in U(N)$. Since we consider a bosonic system, the states are classified by the totally symmetric representations of this $U(N)$ symmetry. In the highest weight notation, they are denoted as $(\lambda,0,\ldots,0)$, and the corresponding Young tableau has one row with $\lambda$ boxes:
\be
\yng(6)\qquad \text{(for $\lambda=6$)}\;.
\ee
The degeneracy  is  the dimension of the representation: 
\be
\mathrm{dim}(\lambda,0,\ldots,0)=\left(\begin{array}{c}N+\lambda-1\\ \lambda \end{array}\right). 
\ee
The thermal partition function of this system is given by
\bea
\calZ(\beta)&=&
\sum_{\lambda=0}^{\infty}\sum_{n_1,\ldots,n_N\ge 0}\delta_{n_1+\cdots+n_N, \lambda}\mathrm{e}^{-\beta E_{\{n_i\}}}\nonumber\\
&=&e^{-\beta N/2} \sum_{\lambda=0}^{\infty}\left(\begin{array}{c}N+\lambda-1\\ \lambda\end{array}\right)
\mathrm{e}^{-\beta \lambda}
\nonumber\\
&=& \frac{1}{\left(2\,\sinh(\beta/2)\right)^N}
\label{partitionf}
\eea
The degeneracy factor grows very rapidly. In fact, in the large-$N$ limit, the states with $\lambda\sim N$ have exponentially large degeneracy.
This is a quite generic feature of thermal partition functions for theories with large global symmetries.

\subsection{Quantum distillation  and graded representations}

We now construct a symmetry-twisted partition function, which is free from  contamination by high-energy states. Define the symmetry operator $\widehat{\Omega}$
\be
\widehat{\Omega}=  \exp\left({\im \sum_{k=1}^N  \alpha_k \widehat Q_k}\right), \quad {\rm where }  \;\;  \alpha_k =  {2\pi\over N}k, \;   \widehat Q_k = \widehat{a}^{\dagger}_k \widehat{a}_k. 
\ee
It acts on the Fock basis as 
\be
|\{n_i\}\rangle \mapsto \exp\left(\im\sum_{k=1}^\infty {2\pi\over N}k\, n_k\right)|\{n_i\}\rangle.
\ee
The corresponding $N\times N$ twist matrix $\Omega_{ij}$ is defined by  
\be
\widehat{\Omega}\widehat{a}^{\dagger}_j \widehat{\Omega}^{-1}= \Omega_{ij}\widehat{a}^{\dagger}_j \quad , 
\ee
and  is diagonal, with entries being powers of the $N$-th root of unity:
\begin{eqnarray}
\Omega=\mathrm{diag}[q,q^2,\ldots,q^{N}] \qquad, \qquad q\equiv \exp\left({2\pi \im\over N}\right)
\label{eq:twist}
\end{eqnarray} 
Using this symmetry operator $\widehat{\Omega}$, we define the symmetry-twisted partition function, 
\be
\calZ_{\Omega}(L)=\mathrm{tr}\left[\widehat{\Omega}\exp(-L \widehat{H})\right]. 
\label{gpf}
\ee
We use the symbol $L$ instead of $\beta$, to further distinguish the twisted partition function (\ref{master}) from the thermal partition function (\ref{eq:thermal}).
Using the Fock basis for the trace, we find that 
\be
\calZ_{\Omega}(L)=e^{-L N/2}\sum_{\lambda=0}^{\infty}\tr_{(\lambda,0,\ldots,0)}(\Omega)\mathrm{e}^{-L \lambda}, 
\label{eq:zho}
\ee
where $\tr_{\cal R}$ means the trace in the representation $\cal R$, labelled in highest weight notation.
 We evaluate the density of states using the following formula for the totally symmetric representation, 
\be
\tr_{(\lambda,0,\ldots,0)}(\Omega)=\sum_{k=0}^{\lambda} {1\over k!} \sum_{r_1,\ldots,r_k\ge 1}\delta_{r_1+\cdots+r_k,\lambda}{\tr_\square(\Omega^{r_1})\cdots \tr_\square(\Omega^{r_k})\over r_1 \cdots r_k}. 
\ee
\begin{figure}[t]
\centering
\includegraphics[scale=0.5]{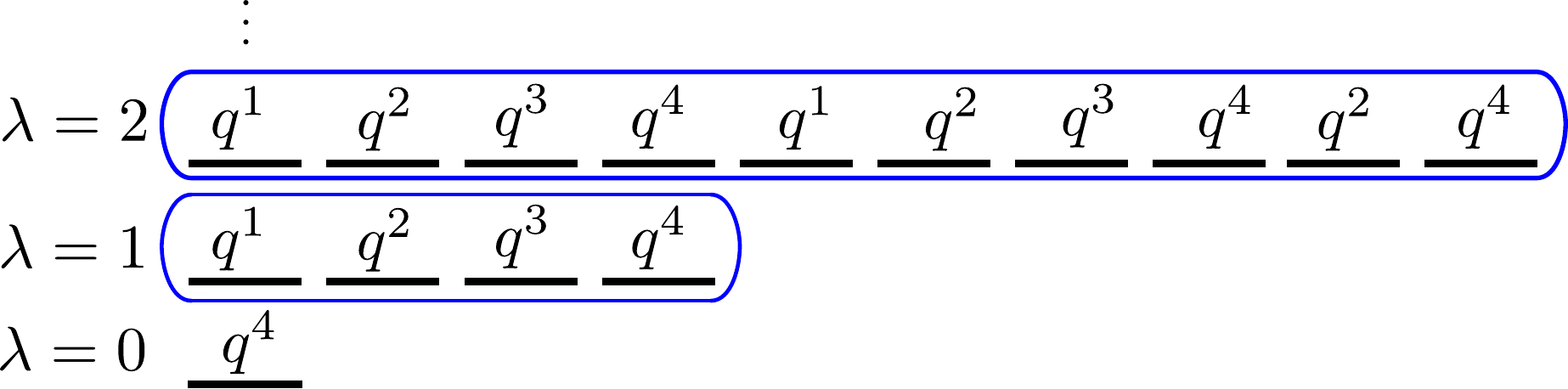}
\caption{Schematic image for the graded partition function (\ref{eq:zho}) for an $N$-dimensional isotropic simple harmonic oscillator, with $N=4$, showing the representations with $\lambda=0, 1, 2$, corresponding to energy $E_{\lambda}=\frac{N}{2}+\lambda$.  Since the phases $q^n=\exp(2\pi i n/4)$ are attached to the states, most of them cancel with one another, and we obtain the information of the ground states.  We will see that a similar structure is also present in certain QFTs.}
\label{fig:graded_trace}
\end{figure}
The trace in the defining representation $(1,0,\ldots,0)=\square$ shows that 
\be
\tr_\square(\Omega^r)=\left\{\begin{array}{cc}
N& \quad(r=0\,\, \mathrm{mod}\; N),\\
0& \quad (r\not=0\,\, \mathrm{mod}\; N). \end{array}\right.
\ee
Using these results, we obtain\footnote{Note the formula: $\sum_{k\ge 0}{1\over k!}\sum_{r_1,\ldots,r_k\ge 1}{\delta_{r_1+\cdots+r_k,\lambda}\over r_1\cdots r_k}=1$, coming from the Taylor expansion of $(1-x)^{-1}=\exp(-\ln (1-x))$.} 
\be
\tr_{(N,0,\ldots,0)}(\Omega^r)=
\left\{\begin{array}{cc}
1& \quad(r=0\,\, \mathrm{mod}\; N),\\
0& \quad (r\not=0\,\, \mathrm{mod}\; N). \end{array}\right.
\label{iden2}
\ee
There is  tremendous cancellation  among degenerate states due to the phase factors attached to them, leading to 
\be
\calZ_{\Omega}(L)=e^{-L N/2} \sum_{\lambda=0}^{\infty}\mathrm{e}^{-L N \lambda}={e^{-L N/2}\over 1-\mathrm{e}^{-N L}}=\frac{1}{2\sinh(L N/2)}. 
\label{1d-N}
\ee
This should be compared with the thermal partition function (\ref{partitionf}). Note the very different behavior in the large scale limit ($\beta$ or $L$, respectively):
\begin{eqnarray}
\calZ(\beta)&= & e^{-\beta N/2}\left(1+N \;e^{-\beta}+\cdots +  \left(\begin{array}{c} 2N-2\\ N-1 \end{array}\right)e^{-(N-1)\beta}  + \left(\begin{array}{c} 2N-1\\ N \end{array}\right)e^{-N \beta}  +    \dots \right) \qquad  \qquad 
\\
\calZ_\Omega(L)&= & e^{-L N/2}\left(1+0  \times e^{-L}+\cdots+ \;\;\;\; 0  \times e^{-(N-1)L} \qquad + 1 \times e^{-N L}+\dots\right). 
\label{eq:zcomp}
\end{eqnarray}
Thus, the symmetry-twisted partition function of the $N$-dimensional isotropic oscillator at radius $L$ maps to the 
thermal partition function of a one-dimensional harmonic oscillator with the inverse temperature $N L$.  Therefore, at large $N$, the symmetry-twisted partition function $\calZ_\Omega(L)$ is much less  contaminated by high-energy states.
The relation to large $N$ volume independence in QFT is discussed in Section \ref{sec:NV}.

This mapping, albeit simple, may in principle have dramatic implications. In particular,  $N$ can be taken to infinity to reach a thermodynamic limit, and the thermal partition function of a thermal system may develop non-analyticities, and associated phase transitions.
But the graded partition function maps to a system with effectively one-degree of freedom, which will not exhibit a thermodynamic phase transition. 
This type of simplification is what we seek in non-trivial QFTs, discussed below in Section \ref{sec:qft}.
This is physically interesting, because if the graded partition function is analytic as a function of $L$, that gives an opportunity to study non-perturbative aspects of asymptotically free  QFTs  precisely at small-$L$ where the theory becomes weakly coupled.  This gives a  Hilbert space reinterpretation of the  phenomena of adiabatic continuity and  semi-classic calculability that have been observed in 2d $\mathbb {CP}^{N-1}$, $O(N)$, principal chiral model, and Grassmannian sigma models~\cite{Dunne:2012ae, Dunne:2012zk,Misumi:2014jua, Misumi:2014bsa, Dunne:2015ywa, Cherman:2013yfa,Cherman:2014ofa,Buividovich:2017jea}, and is also potentially related to  center-stability  and avoidance of Hagedorn instability in 4d QCD(adj) (QCD with multiple adjoint fermions)~\cite{Dunne:2016nmc,Unsal:2007vu, Kovtun:2007py, Unsal:2007jx, Unsal:2008ch, Shifman:2008ja, Shifman:2009tp, Basar:2013sza, Basar:2014jua}. These QFTs have been analyzed from the path integral perspective. Therefore, we next  briefly discuss the  implications of quantum distillation of Hilbert space  in the path integral formulation.

\subsection{Path integral interpretation: Flavor-momentum transmutation}

It is useful to interpret the Hilbert space distillation in the graded partition function $\calZ_{\Omega}(L) $ from the path integral viewpoint.  This  provides a perspective on the volume independence phenomenon in  large-$N$ QFTs. 
In the absence of an operator insertion to the partition function, the Kaluza-Klein modes of a spatially compactified (with radius $L$) QFT are quantized in units of $\frac{2\pi}{L}$, where each KK-mode has $N$-fold degeneracy.  The  insertion of the operator $\widehat \Omega = e^{\im  \sum \alpha_i Q_i} $  in \eqref{gpf} amounts to a finer quantization of the KK-modes,  as the $N$-fold degeneracy of KK-modes is  split   into $N$-levels with 1-fold degeneracy, where the spacing between the levels is now $\frac{2\pi}{LN}$.  In particular, in the $L=$fixed, $N\rightarrow \infty$ limits, the spectrum looks like the perturbative spectrum of a theory on $\R^2$, despite the fact that $L \sim O(N^0)$ can be {\it arbitrarily}  small. We refer to this phenomenon as flavor-momentum transmutation. This is analagous to the color-momentum transmutation in  the  Eguchi-Kawai reductions as discussed by Gross and Kitazawa~\cite{Gross:1982at}, and Gonzalez-Arroyo and Okawa~\cite{GonzalezArroyo:1982hz, GonzalezArroyo:1982ub, GonzalezArroyo:2010ss}.

More formally, in  the coherent state basis, the path-integral expression of the thermal partition function of the $N$-dimensional (bosonic) oscillator is given by 
\be
\calZ (\beta) =\int_{a(\beta)= a(0)}  \Diff a^* \Diff a \exp\left(\int_0^{\beta}\diff \tau\sum_{k=1}^N\left[\im a^*_k \p_{\tau}a_k-a^*_k a_k\right]\right). 
\ee
with periodic boundary conditions on the fields, as appropriate for bosonic degrees of freedom.
The graded partition function can be realized with {\it twisted} boundary conditions on the $a (\tau)$,  
$\calZ_\Omega (L) =\int_{a(L)= \Omega a(0)}  ( \ldots)  $. By a field redefinition, the $a(\tau)$ can be rendered periodic at the price of turning on a particular  $SU(N)$ background field. The result is:  
\be
\calZ_{\Omega}(L)=\int \Diff a^* \Diff a \exp\left(\int_0^{L}\diff \tau\sum_{k=1}^N\left[\im a^*_k\left(\p_{\tau}+\frac{2\pi}{N L}k\right)a_k-a^*_k a_k\right]\right). 
\ee
Decomposing the field with the Matsubara modes, $a_k(\tau)={1\over \sqrt{L}}\sum_{m}\tilde{a}_{k,m}\mathrm{e}^{\im \frac{2\pi}{L}m \tau}$, the exponent becomes 
\bea
&&\sum_{m=-\infty}^{\infty}\sum_{k=1}^N \left[\im \tilde{a}^*_{k,-m}\left(\frac{2\pi}{L} m+\frac{2\pi}{ N L}k\right)\tilde{a}_{k,m}-\tilde{a}^*_{k,-m} \tilde{a}_{k,m}\right]\nonumber\\
&=&\sum_{m'=-\infty}^{\infty}\left[\im \frac{2\pi}{N L}m'\tilde{a}'^*_{-m'}\tilde{a}'_{m'}-\tilde{a}'^*_{-m'} \tilde{a}'_{m'}\right]\nonumber\\
&=&\int_{0}^{N L} \diff \tau' \left[\im a'^*(\tau')\p_{\tau^\prime} a'(\tau')-a'^*(\tau')a'(\tau')\right], 
\eea
where $\tilde{a}'_{m'}=\tilde{a}_{k,Nm}$, with $k=m'$ mod $N$. This defines the partition function of a one-dimensional harmonic oscillator with inverse temperature $N L$. Because of the twisted boundary condition, the flavor-dependent imaginary chemical potential plays the role of refined Matsubara frequencies.  This reproduces exactly \eqref{1d-N}, with the physical implications already discussed after  \eqref{1d-N}.

\subsection{Quantum Distillation in ${\mathbb {CP}}^{N-1}$ Quantum Mechanics} 
As a warm-up to some 2d sigma model QFTs,  we now briefly consider the quantum mechanics of a particle moving on the 
complex projective space, $\mathbb {CP}^{N-1}$:
\be
\mathbb {CP}^{N-1}=\frac{U(N)}{U(N-1)\times U(1)}.
\ee  
This QM problem is exactly solvable, and the group theoretic structure of the corresponding Hilbert space is well known: see for example \cite{Macfarlane2003}.
One may be tempted to think that the Hilbert space again decomposes into representations of $U(N)$, as in the $N$-dimensional simple harmonic oscillator discussed in the previous sub-sections. 
However, the correct global symmetry is in fact $PSU(N)= SU(N)/\mathbb Z_N$.
An analogous comment applies to the familiar example of a QM particle on $S^2$, where we do not have half-integer spin states (in which case the symmetry would be $SU(2)$) but only  integer spin states, so that  the faithful symmetry is $SU(2)/{\mathbb Z}_2=PSU(2)= SO(3)$.

In the   highest weight notation, 
the Hilbert space can be decomposed as \cite{Macfarlane2003}:
\begin{align}
\mathcal{H} \simeq  \bigoplus_{\lambda=0}^{\infty}  \;\;  \mathcal{R}_{\lambda}, 
\end{align}
with ${\cal R}_{\lambda}=(\lambda,\underbrace{0,\ldots,0}_{N-3}, \lambda)$. The degeneracy of the representation  grows extremely quickly:
\be
\dim({\cal R}_{\lambda}) = {2\lambda+N-1\over N-1}\left(\begin{array}{c}\lambda+N-2\\ \lambda \end{array}\right)^2,  
\label{eq:cp-dim}
\ee
Therefore, the  thermal partition function is:
\be
\calZ (\beta)=\sum_{\lambda=0}^{\infty}   \dim({\cal R}_{\lambda}) \mathrm{e}^{-\beta E_{(\lambda,0,\ldots,0,\lambda)}} .
\ee
where the energy eigenvalues are expressed in terms of the quadratic Casimir:
\begin{eqnarray}
E_{(\lambda,0,\ldots,0,\lambda)} =2 \lambda (\lambda+ N-1)
\label{eq:cpn-energy}
\end{eqnarray}

Now consider  the following graded partition function:
\be
\calZ_{\Omega}=\mathrm{tr}\left[\widehat{\Omega}\exp(-L \widehat{H})\right] = \sum_{\lambda=0}^{\infty}\tr_{(\lambda,0,\ldots,0,\lambda)}(\Omega)\mathrm{e}^{-L E_{(\lambda,0,\ldots,0,\lambda)}    },
\label{gpf-2}
\ee
where $\Omega$ is the $U(N)$ twist matrix in (\ref{eq:twist}).  
This requires the knowledge of $\tr_{(\lambda,0,\ldots,0,\lambda)}(\Omega)$, 
which can be found easily using the multiplication identity:
\be
(\lambda,0.\cdots,0,0)\otimes (0,0,\cdots,0,\lambda)=(\lambda-1,0.\cdots,0,0)\otimes (0,0,\cdots,0,\lambda-1)+(\lambda ,0,\cdots,0,\lambda). 
\ee
For example, for  $\lambda=5$ and $N=4$, we have the Young tableaux expression:
\be
\Yvcentermath1
\left(\yng(5,5,5)\otimes \yng(5)\right)
=\left(\yng(4,4,4)\otimes \yng(4)\right) \oplus \yng(10,5,5)
\ee
Therefore, 
\be
\tr_{(\lambda,0,\cdots,0,\lambda)}(\Omega)=\left|\tr_{(\lambda,0,\cdots,0)}(\Omega)\right|^2-\left|\tr_{(\lambda-1,0,\cdots,0)}(\Omega)\right|^2. 
\label{iden}
\ee
This formula has several useful implications. For example, taking $\Omega= {\bf 1}$ gives the dimension of the corresponding representation,  $\tr_{(\lambda,0,\cdots,0,\lambda)}( {\bf 1}) =\dim((\lambda,0,\ldots,0,\lambda))$:
\be
\dim((\lambda,0,\ldots,0,\lambda))=\left(\begin{array}{c}\lambda+N-1\\ \lambda\end{array}\right)^2 -\left(\begin{array}{c}\lambda+N-2\\ \lambda-1\end{array}\right)^2={2\lambda+N-1\over N-1}\left(\begin{array}{c}\lambda+N-2\\ \lambda\end{array}\right)^2,  
\ee
agreeing with the degeneracy (\ref{eq:cp-dim}) of the states in the representation ${\cal R}=(\lambda,0,\ldots,0,\lambda)$.

The expression (\ref{iden}) implies that it is natural to grade with respect to totally symmetric representations (as in the harmonic oscillator example), but of two different sizes. This leads to dramatic cancellations in the graded state sum (\ref{gpf}).
Indeed, using \eqref{iden} and \eqref{iden2}, we find that for the $U(N)$ twist (\ref{eq:twist})
\be
\tr_{(\lambda,0,\ldots,0,\lambda)}(\Omega)=\left\{\begin{array}{cl}
1 & \;\; (\lambda=0 \,\,\, \mathrm{mod}\,\, N), \\
-1 & \;\; (\lambda=1 \,\,\, \mathrm{mod}\,\, N),\\
0 & \;\; (\mbox{others}). 
\end{array}\right.
\label{traces}
\ee
Therefore, the symmetry-twisted partition function reduces  to 
\begin{align}
\calZ_{\Omega}(L)&=  \sum_{k=0}^{\infty} \left( \mathrm{e}^{-L E_{(kN,0,\ldots,0,kN)}}-  \mathrm{e}^{-L E_{(kN+1,0,\ldots,0,kN+1)}}   \right)
\end{align}
In the large-$L$ (arbitrary $N$), as well as large-$N$ (arbitrary $L$) limits,   the sum is dominated by low $k$ values:
\begin{eqnarray}
\calZ_{\Omega}(L)\sim 1-e^{-2 N L} +0+\cdots +0+e^{-4N(N-1/2)L}-e^{-4N(N+1)L}+ \dots
\end{eqnarray}
This should be contrasted with  the thermal partition function:
\begin{eqnarray}
Z(\beta)\sim 1+\left(N^2-1\right)\, e^{-2 N \beta}+\frac{1}{4}N^2(N-1)(N+3) \, e^{-4(N+1)\beta}+\dots
\label{eq:cpn-large-beta}
\end{eqnarray}
We again emphasize that in obtaining $\calZ_{\Omega}(L)$, we did not change the theory. The Hilbert space associated with the graded and thermal partition functions are one and the same. 
In the next section, we generalize this Hilbert space argument to QFT, and explore its connection to mixed anomalies.

\section{Quantum distillation  in  2d QFTs}\label{sec:Grassmannian}
\label{sec:qft}

\subsection{Overview of the 2d bosonic Gr$(N,M)$ sigma model QFT} 
Now, we move to an interesting class of  QFTs, the Grassmannian $\Gr(N,M)$  sigma models in 2d.  This theory has the following properties which makes it interesting in its own right,  in addition to being a testing-ground for ideas in QCD-like theories \cite{Perelomov:1987va,Zakrzewski:1984bt}.  
\begin{itemize}
\item Asymptotic freedom and dynamically induced mass
\item Instantons on any 2 dimensional Euclidean spacetime manifold ${\cal M}_2$, and fractional instantons on $\R \times S^1_L$ with twisted boundary conditions.
\item Confinement in the bosonic model 
\item Interpolation from vector model to matrix model as $M$ interpolates from $O(N^0)$ to $O(N^1)$. 
\item  Quantum distillation in its Hilbert space with a suitably graded (symmetry-twisted) partition function. 

\item Theta angle and a mixed anomaly between  charge-conjugation  symmetry $\mathsf{C}$,  and $PSU(N)$ global symmetry on $\R^2$ at $\Theta=\pi$.  In the supersymmetric 
${\cal N}=(2,2)$ extension, a  mixed anomaly between discrete axial $\mathbb Z_{2N}$ and $PSU(N)$ global symmetry. 
\item    Mixed anomaly  ($\mathsf{C}$-$PSU(N)$) on  $\R^1 \times S^1_L$ at $\Theta=\pi$, provided a unique twisted boundary condition is imposed associated with  
$PSU(N)$ symmetry.\footnote{Note that the anomaly polynomial and  constraints remain the  same as in $\R^2$, despite the fact that  the associated global symmetry is {\it zero-form}. With periodic boundary conditions, the anomaly polynomial does not exist on $\R^1 \times S^1_L$. } 
 In the supersymmetric 
${\cal N}=(2,2)$ extension, a  mixed anomaly ($\mathbb Z_{2N}$-$PSU(N)$), provided the same boundary conditions as in the bosonic case are used for the full super-multiplet.  
 
\end{itemize}
Our goal in this section is to provide the relation between quantum distillation, semi-classical analysis and mixed 't Hooft anomalies. All three concepts are tied with symmetry-twisted boundary conditions in the path integral formulation and its operator image in the Hilbert space interpretation.

\vspace{0.5cm}
\noindent
{\bf Formal aspects:}
We first  consider  purely bosonic sigma models whose  target space is given by the complex Grassmannian manifold $\Gr(N,M)$ \cite{Zakrzewski:1984bt,Perelomov:1987va}. The elementary field  is the mapping:
\begin{eqnarray}
 z(x)\;  : \; {\cal M}_2 \rightarrow  {\rm Gr}(N, M)\equiv  \frac{U(N)}{U(N-M) \times U(M)} 
\label{action-gr}
\end{eqnarray} 
The real dimension of the ${\rm Gr}(N, M)$ space is  equal to the number of 
microscopic degrees of freedom:  
 \begin{eqnarray}
{\rm dim}_{\R}{\rm Gr}(N, M)= N^2-[ (N-M)^2 +M^2] = 2M(N-M)
\label{dim-gr}
\end{eqnarray} 
Note that $M=1$ corresponds to the vector-like theory ${\rm Gr}(N, 1)=\mathbb C \mathbb P^{N-1}$, with dimension $2(N-1)$, while the other extreme of $M$ is the matrix-like theory ${\rm Gr}(2N, N)$, with dimension $2N^2$. Thus the Grassmannian models interpolate between vector-like and matrix-like behavior.
This nonlinear sigma model is realized as a gauged sigma model, in terms of fields $z$ as $N\times M$ complex rectangular matrices, constrained such that $z^\dagger z={\bf 1}_M$.  For that purpose, we first recall that 
\be
{U(N)\over U(N-M)}\simeq \left\{z\in \mathbb{C}^{N\times M}\, \Bigl|\, z^{\dagger}z=\bm{1}_M\right\}. 
\label{eq:coset_representation}
\ee
The action of the $\Gr(N,M)$ model is given by 
\be
S={2\over g^2}\int_{M_2}{\tr}[D(a)z^\dagger \wedge * D(a)z]-\im {\Theta\over 2\pi}\int_{M_2}\tr(F), 
\label{eq:Lagrangian_Gr}
\ee
where $z:{\cal M}_2\to U(N)/U(N-M)$, $a$ is the $U(M)$ gauge field on ${\cal M}_2$, $D(a)=\diff +\im a$ is the covariant derivative, and $F=\im^{-1}D(a)\wedge D(a)=\diff a +\im a\wedge a$ is the $U(M)$ field strength. When it is evident, we simply denote $D=D(a)$. 
The field $z$ transforms under $U(N)\times U(M)$ as $z\mapsto U_N z U_M^{\dagger}$, for $(U_N,U_M)\in U(N)\times U(M)$ where $U(N)$ is global symmetry  and $U(M)$ is gauge structure. The covariant derivative acts on $z$ as 
\be
Dz=\diff z - \im z a,\; Dz^{\dagger}=\diff z^{\dagger}+\im a z^{\dagger}. 
\label{eq:covariant_derivative}
\ee
Since $\tr(F)$ is a $U(1)$ field strength, the topological charge is quantized in the integers:
\begin{align}
Q\equiv {1\over 2\pi}\int_{M_2}\tr(F)\in \mathbb{Z} 
\label{top-charge}
\end{align}
Thus the topological $\Theta$ angle in (\ref{eq:Lagrangian_Gr})  is $2\pi$ periodic, in this field normalization.

An important physical implication of the quantization of the topological charge is the existence of charge-conjugation symmetry $\mathsf{C}$ at $\Theta\in \pi \mathbb{Z}$. Namely, the action (\ref{eq:Lagrangian_Gr}) at $\Theta\in \pi\mathbb{Z}$ is invariant under $\mathsf{C}$ modulo $2\pi$, and the quantum theory is invariant under the charge conjugation. This symmetry will play a role in our discussion of quantum anomalies.

The Grassmannian model has instantons \cite{Zakrzewski:1984bt,Perelomov:1987va}. The Bogomolnyi  factorization of the action gives
\bea
S &=&{2\over g^2}\int \tr[(D z)^\dagger\wedge *D z]\nonumber\\
&=&{1\over g^2}\int \tr\left[(D z \mp \im * D z)^{\dagger}\wedge *(D z \mp \im * D z)\right]\mp\im {2\over g^2}\int \tr(Dz^{\dagger}\wedge D z)\nonumber\\
&=&{1\over g^2}\int \tr\left[(D z \mp \im * D z)^{\dagger}\wedge *(D z \mp \im * D z)\right]\pm {4\pi\over g^2}\int {\tr(F)\over 2\pi}. 
\eea
Thus the BPS/$\bar{\text{BPS}}$ equations are $Dz=\pm \im * D z$, and the action is bounded below, $S\ge {4\pi\over g^2}|Q|$, for each topological sector of charge $Q$, with equality satisfied on BPS solutions.

The instanton action $S_{\cal I}$,  instanton amplitude, $\Theta$-angle, dynamical strong scale $\Lambda$, and $\beta$-function are related as: 
\begin{align}
S_{\cal I} &= \frac{  4 \pi }{g^2}, \qquad {\cal I}_{2d} \sim e^{- \frac{  4 \pi }{g^2(\mu)} + i \Theta},   
\qquad  \Lambda^{ \beta_0} = \mu^{ \beta_0} {\cal I}_{2d}, \qquad {\rm any} \; (N, M)
\label{instanton}
\end{align}
where $\beta_0 =N$ is the leading order $\beta$-function \cite{Morozov:1984ad}. In particular, the strong scale is related to a {\it fractional power} of the instanton amplitude, $  {\cal I}_{2d}^{1/\beta_0} \mu= \Lambda$. Such effects play an important role in the quantum Grassmannian field theory with  symmetry-twisted compactification \cite{Dunne:2015ywa}.

\vspace{0.5cm}
\noindent
{\bf Spectrum and representation:}  Let us now consider the spectrum of massive particles. As an analogy with QCD, one may consider 
the $z$ field as a ``quark", and $z^{\dagger}$ as an ``anti-quark". The gauge structure of the theory is $U(M)=(SU(M)\times U(1) )/\mathbb{Z}_M$, and the global symmetry is  $PSU(N)$. The fields $z$  and $z^{\dagger}$  would transform as fundamental $N$ and anti-fundamental $\bar N$ under the global symmetry, but these fields have both global and gauge index, and in particular, they are not gauge invariant. More precisely, 
\begin{align}
z \in \left(\yng(1)_N,  \overline {\yng(1)}_M \right) 
\end{align} 
However, the  bound states of $z$  and $z^{\dagger}$ are gauge invariant. The leading gauge invariant operators in the theory are: 
\begin{align}
 (z z^{\dagger})_{j}^k(x) \in  \left({\rm Adj}_N,  1_M \right)    , \qquad    z (x) (e^{i \int_x^y a}) (z^{\dagger}) (y)  \in  \left({\rm 1}_N,  1_M \right) 
 \label{states-2}
\end{align} 
which are in the adjoint and singlet representations, respectively.  In the large-$N$ limit, with $M$-finite,  the adjoint and singlet representations become degenerate \cite{Witten:1978bc}.

The fact that physical states in the spectrum transform in the adjoint or product of adjoints and singlet representation (instead of the fundamental and other representations) follows from the fact that the global symmetry of the theory is not $SU(N)$, but is $PSU(N)= SU(N)/\mathbb Z_N$.  In this sense, the symmetry with the faithful representations is  $PSU(N)$, and the same thing occurs in our ${\mathbb CP}^{N-1}$ QM example in Section \ref{sec:qm}.

\subsection{Hilbert space interpretation of twisted boundary conditions}
The theory has a vacuum state, which can be described in detail semi-classically \cite{Dunne:2012ae,Dunne:2012zk}.  The excited states,  as in the quantum mechanical example, lie in representations of  $PSU(N)$, and the Hilbert space  admits a corresponding decomposition.
The analysis is similar to the quantum mechanical ${\mathbb CP}^{N-1}$ example, except that, unlike in QM where excited  singlet  states do not appear in the spectrum, now one has singlet higher energy states as well, due to the existence of operators as in \eqref{states-2}.   
Interestingly, and to a certain extent even more dramatically than in the QM examples, in the large-$N$ limit this implies that the graded state-sum in the Grassmannian QFT is even simpler. The Hilbert space construction  in this subsection follows Sulejmanpasic \cite{Sulejmanpasic:2016llc}.
 Consider the insertion of the following operator into the trace:
\be
\widehat{\Omega}=  e^{\im \sum_k  \alpha_k \widehat Q_k}  
\ee
Using this symmetry operator $\widehat{\Omega}$, we define the symmetry-twisted partition function, 
\be
\calZ_{\Omega}(L)=\mathrm{tr}\left[\widehat{\Omega}\exp(-L \widehat{H})\right]. 
\label{gpf-qft}
\ee
The first excited states of the system are the adjoint and the singlet: 
\begin{align}
 &\text{adjoint}: \quad (1,0,\ldots,0,1), \qquad  {\rm degeneracy}= \;  N^2-1 \cr 
 &\text{singlet}: \quad (0,0,\ldots,0,0), \qquad  {\rm degeneracy}= \;  1
 \end{align}
The key facts are  \eqref{traces}  and  that the energy difference between these two states is $O\left(\frac{1}{N}\right)$ in the large-$N$ limit.  
This leads to a  striking difference between the  thermal state sum and  the graded state sum, since these low-lying states contribute as 
\begin{align}
& \calZ (\beta)  \approx  + (N^2-1) \times e^{-\beta E_{\rm adj}}   + 1 \times  
e^{-\beta E_{\rm  singlet}}  +\dots \underbrace{\rightarrow}_{N   \rightarrow   \infty}     
N^2 \times e^{-\beta E_{\rm adj}}   \cr
&\calZ_\Omega (L)  \approx ( -1)  \times e^{-\beta E_{\rm adj}}    \qquad  +  1  \times
e^{-\beta E_{\rm singlet}}  +\dots \underbrace{\rightarrow}_{N   \rightarrow   \infty}     0
 \end{align}
Thus the important effect of the grading is to turn the $(N^2-1) e^{-\beta E_{\rm adj}}$ contribution into a contribution $(-1) e^{-\beta E_{\rm adj}}$, because of \eqref{traces}. This furthermore produces a relative sign between the two leading contributions, $-1 \times e^{-\beta E_{\rm adj}}$ and $+1\times e^{-\beta E_{\rm singlet}}$, in $\calZ_\Omega (L)$, which leads therefore to a cancellation in the large-$N$ limit, due to the degeneracy of $E_{\rm adj}$ and $E_{\rm singlet}$ at large $N$.  For higher energy states, for any $k$-index representation with $k\not = 0$ modulo $N$, $\tr_{\cal R}[\Omega]=0$.  And the next states that can potentially contribute have masses of order $O(N)$. 
  Therefore, we find that the graded partition function has a simple large $N$ limit:
\be
\lim_{N \rightarrow \infty}   \calZ_{\Omega}(L)  \sim \mathrm{e}^{-L E_{(0,0,\ldots,0,0)}}
\label{simple2}
\ee
In particular, it is analytic as a function of $L$, and cannot exhibit a phase transition as $L$ changes. 

Note that analyticity of  \eqref{simple2}  is in sharpe contrast with the behavior of  thermal partition function,  which 
has a deconfinement temperature $T_d= \frac{\Lambda}{\ln N} $, as a result of which the theory becomes deconfined at large-$N$ limit  once a temperature is turned on  \cite{Affleck:1979gy}.   In distinction, \eqref{simple2} is analytic as a function of $L$, and is dominated by the same state both at small and large $L$.  In other words,  $\beta_{\rm critical} \rightarrow \infty$  in Affleck's study   \cite{Affleck:1979gy},  while  $L_c \rightarrow 0$  in \cite{Dunne:2012ae}.

\subsection{Mixed anomalies on $\R^2$}
In this section, we consider the constraint on the vacuum structure coming out of the 't Hooft anomaly matching. Following Refs.~\cite{Tanizaki:2017qhf, Tanizaki:2017mtm}, we construct the 't Hooft anomaly under  adiabatic $S^1$ compactification starting from that of the Grassmannian sigma model on $\R^2$.  
To describe the mixed anomalies, we consider two examples. 
\begin{itemize} 
\item Bosonic $\Gr(N, M)$ model  at $\Theta=\pi$.  The mixed 't Hooft anomaly\footnote{It has recently been shown that the bosonic $\mathbb{CP}^1$ model also has a mixed 't Hooft anomaly between $\mathsf{C}$, parity $\mathsf{P}$, and time reversal $\mathsf{T}$ symmetries~\cite{Sulejmanpasic:2018upi} (see also \cite{Metlitski:2017fmd}). In this paper, we do not discuss this anomaly, but it is an interesting future question whether this anomaly can also persist under a certain boundary condition. } is between $PSU(N)$ and charge conjugation symmetry $\mathsf{C}$. 
\item Supersymmetric ${\cal N}=(2,2)$ $\Gr(N, M)$ model.  The mixed  anomaly is between $PSU(N)$ and axial $\mathbb Z_{2N}$ symmetry. 
\end{itemize} 

The reason for this exercise is two-fold. First, the anomaly has the important effect that it dramatically reduces the possibilities of IR theories. Second, despite the fact that we are considering zero-form symmetries, the anomaly polynomial persists upon compactification on $\R \times S^1_L$,  if and only if  one uses a {\it unique}  twisted boundary condition,  corresponding precisely to the graded partition function $\calZ_{\Omega}(L)=\mathrm{tr}\left[\widehat{\Omega}\exp(-L \widehat{H})\right]$. In path integral language, this corresponds to a background $SU(N)$ holonomy on the compactified spacetime manifold, $\R \times S^1_L$, as used in the semi-classical studies of these theories 
on $\R \times S^1$ \cite{Dunne:2012ae,Dunne:2012zk,Dunne:2015ywa}.
In other words, there is a precise relation between the  persistence of the anomaly upon compactification, the Hilbert space quantum distillation, and the semi-classical analysis of the corresponding path integrals. We first describe the mixed anomaly, and then discuss some of the physics associated with it.

\subsubsection{Mixed anomaly for bosonic $\mathrm{Gr}(N,M)$ model at $\Theta=\pi$}
We first gauge the $PSU(N)$ symmetry, and show that one loses  $2\pi$ periodicity of the $\Theta$ angle. As a result,  $\mathsf{C}$ will be  lost as a symmetry, implying a mixed anomaly. 
 
To gauge the flavor symmetry, we introduce a background $SU(N)$ one-form gauge field $A$, and a $\mathbb{Z}_N$ two-form gauge field  $B$. The discrete two-form gauge structure can be realized  as a pair $(B, C)$,  a $U(1)$ two-form gauge field $B$ and a $U(1)$ one-form gauge field $C$ satisfying the constraint $NB+\diff C=0$. 
 
Minimal coupling  gives the action:
\be
S_{\mathrm{gauged}}={2\over g^2}\int_{\R^2}{\tr}[D(a,\widetilde{A})z^{\dagger}\wedge *D(a,\widetilde{A})z]-\im {\Theta\over 2\pi}\int_{\R^2}[\tr (F)+B],
\ee
Here, $D(a,\widetilde{A})z=\diff z -\im z a + \im \widetilde{A}z$, and $\widetilde{A}=A+{1\over N}C$ is the $U(N)$ gauge field. To obtain this action, we impose the manifest $U(1)$ one-form gauge invariance under 
\bea
B\mapsto B-\diff \lambda,\;\; \widetilde{A}\mapsto \widetilde{A}+\lambda, \;\; a\mapsto a+\lambda. 
\eea
By checking the spectrum of gauge-invariant  line operators, we can immediately see that the faithful symmetry group $SU(N)/\mathbb{Z}_N$ is correctly gauged~\cite{Kapustin:2014gua}: If we consider the $U(N)$ Wilson line in the fundamental representation, $W(C)=\mathrm{tr}\mathcal{P}\exp(\im \int_C \widetilde{A})$, then $W(C)^k$ is not one-form gauge invariant for any $k\not=0$ mod $N$. To maintain  gauge invariance, we must attach the surface operator $\exp(-\im k \int_S B)$ with $\p S=C$, but it depends on the topology of the surface and they are no longer genuine line operators. 

A direct consequence at this stage is that  the topological term \eqref{top-charge}  in the partition function is no longer quantized in integer units. Instead,  
\begin{align}
Q=  {1 \over 2\pi}\int_{\R^2}[\tr (F+B)] = \underbrace{ \int_{\R^2}  \frac{\diff(\tr(a))}{2\pi}}_{\in \mathbb Z} +
 \underbrace{{M \over  N} \int_{\R^2}  \frac{\diff C}{2\pi}}_{\in \frac{M}{N} \mathbb Z}  \in \frac{\mathrm{gcd}(N,M)}{N} \mathbb Z
\label{non-integer}
\end{align} 
which implies that the $\Theta$ angle is no longer $2\pi$ periodic, but is  $2\pi N/{\rm gcd}(M, N)$ periodic. Since strict   $2\pi$ periodicity was crucial for 
 $\mathsf{C}$ symmetry at $\Theta=\pi$, now, this symmetry is lost, implying a mixed anomaly. Before discussing the implications of this mixed anomaly, let us examine the same effect from the perspective of the partition function.

Denote the partition function in the background $(A, B)$ gauge fields as 
\be
\calZ_{\Theta}[(A,B)]=\int \Diff z^{\dagger}\Diff z \Diff a \exp(-S_{\mathrm{gauged}}). 
\ee
where, in the path integral, we integrate only over the fields that are present in the microscopic theory. 
Applying the charge-conjugation operation to this partition function, one observes that   $\calZ$ at $\Theta=\pi$ is changed  as:
\be
\calZ_{\pi}[\mathsf{C}\cdot(A,B)]=\calZ_{\pi}[(A,B)] \exp\left(-\im M  \int_{\R^2}  B\right)
\ee
in the presence of the  background gauge field.  This is of course the same effect that rendered the topological charge non-integer in \eqref{non-integer}. This gives the 't Hooft anomaly, or global inconsistency. 

To demonstrate that this is a genuine anomaly, we must show that it is irremovable by a local counter-term in  two dimensions. 
Gauge invariance says that such a term is given by
\be
\im k \int B, 
\ee
where $k=0,1,\ldots, N-1$. When this term is added, the $\mathsf{C}$ invariance at $\Theta=\pi$ requires 
\be
2k-M=0 \qquad \text{modulo}\quad N\quad .
\ee
Therefore, the apparent anomaly can be removed if and only if $2k = M$ (mod $N$).  
 Doing the same analysis at $\Theta=0$, we get $2k=0$ mod $N$, and there is no anomaly at $\Theta=0$, e.g., by choosing $k=0$.

There are four cases. If $(N, M)$= (even, odd), there is no solution for $k$ at $\Theta=\pi$, and there is an anomaly. 
If $(N, M)$= (even, even), the solution at $\Theta=0$ is $k=0$, and  at $\Theta=\pi$ is $k=M/2$ and there is no counter-term that is good at both $\Theta=0, \pi$.  This is the global inconsistency condition~\cite{Gaiotto:2017yup, Tanizaki:2017bam, Kikuchi:2017pcp}.  Below, we assume the vacuum at $\Theta=0$ is trivial  in such cases;  hence the vacuum at $\Theta=\pi $  cannot be trivial. There is a similar discussion for  $(N, M)=$ (odd, even) and  $(N, M)=$ (odd, odd).

As a result of the anomaly, there are three possible IR realizations: 
\begin{enumerate}
\item  $\mathsf{C}\times PSU(N)$ ought to be spontaneously broken to an anomaly-free subgroup, 
\item There is a TQFT in the IR. A TQFT implies that there cannot be a trivial (unique) ground state  once this theory is considered on arbitrary manifolds. 
\item There is a CFT at the critical point. 
\end{enumerate}
Because of the 2d nature of the theory,  the Mermin-Wagner-Coleman theorem~\cite{mermin1966absence, Coleman:1973ci} implies that the continuous global symmetry   $PSU(N)$ cannot be broken. As a result, if the second and third options do not take place\footnote{For  spin chains, it has been  shown that topological order does not appear in the IR~\cite{PhysRevB.83.035107}, and thus the topological order can be ruled out when the field theory appears as an IR description of certain spin systems. On the other hand, a CFT is possible, and indeed this is a part of the Haldane conjecture for half-integer antiferromagnetic spin chains~\cite{Haldane:1982rj, Haldane:1983ru}. Since they correspond to the $\mathbb{CP}^1$ model at $\Theta=\pi$, CFT matches the anomaly when $N=2$ and $M=1$. }, then the charge conjugation $\mathsf{C}$ must be spontaneously broken.

Before discussing what happens to these anomaly considerations upon compactification of the QFT's spacetime manifold, we also describe our second example, the supersymmetric  ${\cal N}=(2,2)$ $\Gr(N, M)$.

\subsubsection{Mixed anomaly  for supersymmetric   $\mathrm{Gr}(N,M)$ sigma model}

The bosonic Grassmannian theory  can be extended to an  ${\cal N}=(2,2)$ supersymmetric model by introducing a 
 Dirac Fermi field $\psi$, the fermionic partner of the $z$ field, constrained to satisfy  $z^{\dagger} \psi=0_M$.
The supersymmetric theory, apart from the $PSU(N)$ symmetry,  also possesses a $U(1)_A$  symmetry at the classical level, transforming the Dirac  fermion 
  $\psi= \left( \begin{array}{l}  \psi_{+} \cr 
\bar \psi_{-} 
\end{array}  \right)
  $  
as 
\begin{equation}
U(1)_A : \psi 
 \mapsto e^{ i \sigma_3 \alpha} \psi   
\label{eq:U1axial}
\end{equation} 
where  $\psi_{\pm}$  are the right/left  movers. Under this transformation, the fermionic measure changes as 
\be
\Diff \overline{\psi}\Diff \psi\mapsto \Diff \overline{\psi}\Diff \psi \exp\left(-2\alpha\,\im {N\over 2\pi}\int_{\R^2}\tr(\diff a +\im a \wedge a)\right). 
\ee
The $U(1)_A$ symmetry has a global anomaly and is reduced to ${\mathbb Z}_{2N}$ quantum mechanically.  
To see this, note that the axial charge conservation can be written as 
\begin{align}
\Delta Q_A= (2N) \times {1\over 2\pi}\int_{\R^2}\tr(F)  \in 2N  \mathbb Z 
\end{align}
as a result of  integer-quantization of topological charge. Thus, (\ref{eq:U1axial}) is a symmetry only when $\alpha\in {2\pi\over 2N}\mathbb{Z}$. Relatedly, note that the 2d instanton amplitude is 
\begin{equation}
I_{\rm 2d} =e^{-S_I}   ( \tr \psi_{-}   \psi_{+ })^N     
\end{equation} 
and the global symmetry is just  ${\mathbb Z}_{2N}$.  

Following the same analysis as in the bosonic discussion, there is a 
mixed anomaly  between the $PSU(N)$ and ${\mathbb Z}_{2N}$ symmetries. To show this, gauge the  $PSU(N)$ symmetry by introducing a $U(N)$ gauge field $\widetilde{A}$ and a $U(1)$ two-form gauge field $B$ with the constraint $N B+\diff (\tr(\widetilde{A}))=0$. Under the $U(1)_A$ transformation (\ref{eq:U1axial}), the fermion measure is changed by 
\bea
\Diff \overline{\psi}\Diff \psi&\mapsto& \Diff \overline{\psi}\Diff \psi \exp\left[2\alpha\,\im \left({M\over 2\pi}\int_{M_2}\tr(\diff \widetilde{A} +\im \widetilde{A} \wedge \widetilde{A})-{N\over 2\pi}\int_{M_2}\tr(\diff a +\im a \wedge a)\right)\right] \nonumber\\
&=&\Diff \overline{\psi}\Diff \psi \exp\left[-2N\alpha\,\im \left({1\over 2\pi}\int_{M_2}\tr(\diff a +\im a \wedge a)+{M\over 2\pi}\int_{M_2}B \right)\right]
\eea
As a result, the  axial charge non-conservation becomes 
 \begin{align}
 \Delta Q_A\in 2M \mathbb{Z}+2N\mathbb{Z}=2\mathrm{gcd}(N,M)\mathbb{Z}. 
\end{align}
 This shows that once $PSU(N)$ is gauged,  ${\mathbb Z}_{2N}$ is no longer a symmetry,  reducing it all the way down to ${\mathbb Z}_{2 \gcd(M, N)}$: Under the $\mathbb{Z}_{2N}$ transformation, 
\be
\calZ[(A,B)]\mapsto \calZ[(A,B)] \exp\left(-\im M \int_{\R^2}B\right), 
\ee  
and we obtain the 't Hooft anomaly.

Similar to the bosonic model, since the continuous global symmetry cannot be broken (invoking the Mermin-Wagner-Coleman theorem), the only option in the IR is spontaneous breaking of $\mathbb Z_{2N}$ down to $\mathbb Z_{2\gcd(M,N)}$ or smaller subgroups, assuming a TQFT or a CFT does not arise. 

Indeed, the first option is consistent 
 with dynamical breaking of chiral symmetry 
by a fermion-bilinear condensate: 
 \begin{equation}
  \label{cc1}
 \langle  \tr  \psi_{-}   \psi_{+ }  \rangle = N \Lambda\, e^{i \frac{2 \pi k}{N}}, \qquad  k=0,1, \ldots, N-1 
 \end{equation}
  leading to $N$ isolated vacua, as well as being consistent with the supersymmetric index $I_W= \tr((-1)^F)=N$ \cite{Witten:1982df,Witten:1982im}.

\subsection{Adiabatic  compactification and persistence of anomaly polynomial}
The graded partition function in the operator formalism \eqref{gpf}, translated   to  the path integral formulation, corresponds 
to considering a path integral with  symmetry twisted boundary conditions along the $S^1_L$ compactification 
\be
z(x^1, x^2+L)= \Omega z(x^1, x^2), 
\label{bc}
\ee
where 
\be
\Omega = \mathrm{diag}(1,\omega, \ldots, \omega^{N-1}) \in U(N). 
\label{holonomy}
\ee
Equivalently, one may consider the insertion of a background holonomy $\Omega$ into the action. 
Under this boundary condition, we now show that  the two-dimensional 't Hooft anomaly survives even after circle compactification on 
$\R \times S^1_L $ at arbitrarily size $L$ of the circle $S^1_L$. 

The key observation is that the boundary conditions \eqref{bc}, or equivalently, the background holonomy \eqref{holonomy}, remain invariant under the intertwined combination of a center-transformation with phase $\omega$ and a $\mathbb Z_N$ cyclic permutation~\cite{Tanizaki:2017qhf}.  Neither of these individual transformations leave the boundary condition invariant, however, combining the center transformation  with the opposite cyclic permutation leaves the boundary condition intact. 

When taking this symmetry-twisted boundary condition, we have the $\mathbb{Z}_N$ symmetry generated by the following transformation, 
\be
z\mapsto S z,\,\, \Phi\mapsto \omega^{-1} \Phi, 
\ee
where $\Phi=\mathcal{P}\exp(\im \int_{S^1} a)$ is the $U(M)$ Polyakov loop, and $S$ is the shift matrix of $SU(N)$ labels:
\be
S=\left(\begin{array}{ccccc}
0&1&0&\cdots&0\\
0&0&1&\cdots&0\\
 \vdots&\vdots &\vdots & &\vdots\\
0&0&0&\cdots&1\\
1&0&0&\cdots&0
\end{array}\right).
\ee
We must combine these two transformations in order to keep  the boundary condition invariant.\footnote{This is parallel to   the emergence of the color-flavor center symmetry in $SU(N_f)$-symmetric QCD with $N_f$ flavor of massive fermions with ${\rm gcd}(N_f, N)   \geq 2$~\cite{Cherman:2017tey, Tanizaki:2017qhf, Tanizaki:2017mtm}. }

Since the $\mathbb{Z}_N$ transformation involves the transformation on the Polyakov loop, we must introduce the $\mathbb{Z}_N$ two-form gauge field, 
\be
B=B^{(1)}\wedge L^{-1}\diff x^2,
\ee
when gauging this $\mathbb{Z}_N$ symmetry. Here, $B^{(1)}$ is a $\mathbb{Z}_N$ one-form gauge field. 
As a result, we obtain 
\be
\calZ_{\pi}[\mathsf{C}\cdot B^{(1)}]=\calZ_{\pi}[B^{(1)}]\exp\left(-\im M \int B^{(1)}\right). 
\ee
This gives the 't Hooft anomaly or global inconsistency condition on $\mathbb{R}\times S^1_L$ under the symmetry-twisted boundary condition, with the same implications as those discussed for the theory on $\R^2$. 

An almost identical consideration also shows that in the ${\cal N}=(2,2)$ supersymmetric  Gr$(N, M)$ model, the anomaly remains essentially the same between  $ PSU(N)$ and $\mathbb Z_{2N} $, the discrete axial symmetry, upon compactification of the theory on $\R^1 \times S^1_L$.

\subsection{Overview of small-$L$ resurgent semiclassics} 
The non-perturbative dynamics of the $\mathbb {CP}^{N-1}$ and Gr$(N, M)$  models with $\Omega$-twisted boundary condition has been examined in detail in  the recent literature  ~\cite{Dunne:2012ae, Dunne:2012zk,Misumi:2014jua, Misumi:2014bsa, Dunne:2015ywa}.  
Here we highlight some of the connections with the analysis of this current paper. 

As discussed earlier, Gr$(N, M)$ admits instanton solutions with topological charge  $Q_T=1$. In the presence of the $\Omega$-background, 
\eqref{instanton}, these 2d instantons split into $N$ minimal action 
kink-instantons, each with topological charge $Q_T= \frac{1}{N}$ \cite{Tong:2002hi, Bruckmann:2007zh, Brendel:2009mp,Dunne:2012ae}. Note that in the study of Affleck, where  $\Omega=1$, fractionalization to 
$N$ kink-instanton does not take place~\cite{Affleck:1979gy}. The fractionalization is similar to 4d instantons splitting into  $N$ monopole instantons in the presence of a non-trivial holonomy in gauge theories on $\R^3 \times S^1_L$. 
The kink-instanton and anti-instanton  amplitudes are given by 
 \begin{align}
& {\cal K}_{j,k}    = e^{-S_I/N} e^{ i \frac{\Theta + 2 \pi k}{ N} } = e^{ -\frac{4 \pi}{g^2 N} }e^{ i \frac{\Theta + 2 \pi k}{ N} }
, \qquad     \overline {\cal K}_{j,k} =  {\cal K}_{j,k}^{*} 
 \end{align}
 for each $j=0, 1, \ldots, N-1$, and $k=1, \ldots, N$ is a branch label. 
    As asserted, the  2d  instanton  amplitude is a composite of the $N$-kink instantons:\footnote{A similar splitting formula for saddle contributions applies for other 2d sigma models, such as the principal chiral model and the $O(N)$ model with $N> 3$; note that these are models without instantons \cite{Cherman:2013yfa,Cherman:2014ofa,Dunne:2015ywa}.}
  \begin{align}
 {\cal I} =\prod_{j=0}^{N-1}    {\cal K}_{j,k}   =  e^{-S_I} e^{ i \Theta } = e^{ -\frac{4 \pi}{g^2} + i \Theta}
  \end{align}
For example, the mass gap in the system is generated by the proliferation of kink-instantons, and is given by \cite{Dunne:2016nmc,Dunne:2012zk}
\begin{align}
 m_g(\Theta)    \propto   \max_k  \left( {\cal K}_{j,k}   + \overline {\cal K}_{j,k}   \right) 
 = \Lambda  \max_k    \cos{ \left( \frac{\Theta + 2 \pi k}{ N} \right) },   \qquad \Lambda = L^{-1} e^{ -\frac{4 \pi}{N g^2(L^{-1})} }
  \label{gap}
\end{align} 
This formula is rather intriguing. It shows that 
\begin{itemize}
\item
The mass gap may be induced  by semi-classical instantons with action  
$S_{\cal K} = S_{\cal I}/N$.   
\item
These semi-classical effects survive the large-$N$ limit, $e^{ -\frac{4 \pi}{g^2N }} \sim O(N^0)$, unlike the 2d instanton, which is suppressed as   $e^{ -\frac{4 \pi}{g^2 }} \sim e^{-N}$.  
 \item
 The result is multi-branched, and the choice associated with the vacuum of the theory  is non-analytic at  $\Theta=\pi$, related to spontaneous $\mathsf{C}$-breaking and the existence of two vacua.   
\end{itemize}

The scenario with two vacua that we find in the semi-classical domain is one of the  possible outcomes of the anomaly consideration for these models at $\Theta=\pi$. It is also interesting to see that for $N=2, M=1$, namely the $\mathbb {CP}^1$ model,  the mass gap vanishes at $\Theta=\pi$, at leading order in semi-classics. 
 On $\mathbb{R}^2$, this is a part of the Haldane conjecture~\cite{Haldane:1982rj, Haldane:1983ru, Affleck:1986pq, Affleck:1987vf, Haldane:1988zza}. 
This is consistent with the CFT possibility arising from the 't Hooft anomaly matching argument. 

In Section \ref{why}, we explain the relation between the semi-classical analysis, and why it works the way it does on $\R\times S^1_L$ while also producing some of the non-perturbative aspects  of the theory on $\R^2$,  from the point of view of the Hilbert space quantum distillation and mixed anomalies.

\subsection{Large-$N$  volume independence and flavor-momentum transmutation}
\label{sec:NV}

Turning on the $\Omega$-background ensures that the anomaly polynomial survives upon compactification. There is another important property associated with this background. In the large-$N$ limit, $SU(N)$ singlet observables as well as their correlation functions become independent of $N$ for any finite value of $L$. 
Let us explicitly show it by using the technique developed by Sulejmanpasic~\cite{Sulejmanpasic:2016llc}. 

The perturbative intuition behind volume independence is the following. If one imposes a trivial $\Omega=1$ background, then Kaluza-Klein momenta are naturally quantized in units of $\frac{2\pi}{L}$.  In the infinite volume limit, $L\rightarrow  \infty$, the momentum modes become a continuum, producing perturbation theory on $\R^2$. However, with the non-trivial twist-$\Omega$, the momentum modes are quantized in a much finer spectrum, in units of $\frac{2\pi}{LN}$. Thus the flavor background $\Omega$ transmutes into fractionalized momenta. In this case, to obtain perturbation theory on $\R^2$, at least, the planar perturbation theory, there are two options. {\it i)} $N=$fixed, $L\rightarrow \infty$ as before or 
{\it ii)}   $L=$fixed, $N\rightarrow \infty$. In this latter case, one can  derive global symmetry singlet observables by using reduced QM system. For example,  the renormalization group $\beta$-function  
of QFT can be derived from reduced QM, 
by using  a similar construction of the matrix model derivation of the $\beta$ function of Yang-Mills  by Gross and Kitazawa~\cite{Gross:1982at}. 

Technically, the construction works as follows (see \cite{Sulejmanpasic:2016llc} for the $\mathbb C\mathbb P^{N-1}$ argument). 
We fix $M$ and take the large-$N$ limit of the $\Gr(N,M)$ model. For simplicity, we take $\Theta=0$, but it can easily be incorporated. 
To take the large-$N$ limit, we define the 't Hooft coupling by 
\be
2f_0=g^2 N. 
\ee
The action becomes
\be
S_{\rm eff}={N\over f_0}\int_{{\cal M}_2} \diff^2 x \, \tr\left\{D_{\mu}z^{\dagger} D_{\mu}z+\im \lambda(z^{\dagger}z-1)\right\}, 
\ee
where we introduced the Lagrange multiplier field $\lambda$, which is a real $M\times M$ matrix-valued field. 

We consider the circle compactification ${\cal M}_2=\R \times S^1_L$, and we take the symmetry twisted boundary condition along $S^1_L$. Integrating out $z$ and $z^{\dagger}$, we obtain the effective action 
\be
S_{\mathrm{eff}} (a_\mu, \lambda)=-\mathrm{Tr}\ln[-D_{\mu}^2+\im \lambda]-{\im N\over f_0}\int \diff^2 x\, \tr(\lambda). 
\ee
We evaluate the functional trace using a plane wave basis:
\bea
&&\mathrm{Tr}\ln[-D_{\mu}^2+\im \lambda]\nonumber\\
&=&\int \diff^2 x\left(\sum_{s=0}^{N-1}{1\over L}\sum_{n\in\mathbb{Z}}\int{\diff k_1\over 2\pi}\tr\ln \left[-\left(D_1+\im k_1\right)^2-\left(D_2+\im {2\pi n+(2\pi s/N)\over L}\right)^2+\im \lambda\right]\right)\nonumber\\
&=&N\int \diff^2 x\left({1\over N L}\sum_{m\in\mathbb{Z}}\int{\diff k_1\over 2\pi}\tr\ln \left[-\left(D_1+\im k_1\right)^2-\left(D_2+\im {2\pi m\over N L}\right)^2+\im \lambda\right]\right). 
\eea
where in the intermediate step, the sum over Kaluza-Klein modes and flavor modes  merges to a much finer sum. As a result of this transmutation, 
in the large-$N$ limit, the explicit $L$ dependence in the integrand of $\int \diff^2 x$ disappears~\cite{Sulejmanpasic:2016llc}:
\be
\lim_{N\to \infty}(...)=\int{\diff^2 k\over (2\pi)^2}\tr\ln\left[-(D_{\mu}+\im k_{\mu})^2+\im \lambda\right]. 
\ee
Therefore, we obtain the effective action as $S_{\rm eff}= N s_{\rm eff}(a_\mu, \lambda)$,  where  $s_{\rm eff}(a_\mu, \lambda)$ is $N$-independent, and hence the action is  suitable for  saddle point analysis in the $N\to \infty$ limit. The effective action is identical to the one on $\R^2$ except that in the  $\int_{\R^2} $ integral, the domain is now
${\cal M}_2=\R \times S^1_L$. This is immaterial for the saddle point analysis. Studying the gap equation in the large-$N$  analysis, one finds 
$a_{\mu}=0, \lambda=m_g^2$~\cite{Witten:1978bc}. Here, $m_g=\mu e^{- \frac{  4 \pi }{g^2(\mu)N}} $ has an interpretation as the mass of $z$ quanta, now derived from reduced model.   This argument proves the large-$N$ volume independence in the $\Omega$-background. 

The analysis is valid for any $SU(N)$ invariant operators, and thus the large-$N$ scaling of correlation functions of the topological charge can also be obtained. This recovers the $\Theta$ dependence of the vacuum for large $N$, as well as the adiabatic continuity of the vacuum structure. 


\section{Conclusions: What is happening and why  is it  happening?}
\label{why}
The interpretation of what is going on both in the Hilbert space and path integral formulation of these theories is actually quite intriguing. 
Consider the following graded state-sums, and their path integral realizations:
\begin{align} 
\calZ_{\Omega}(L) & =\mathrm{tr}\left[\widehat{\Omega}\exp(-L \widehat{H})\right] \cr
&=  \int_{z(x^1, x^2+L)= \Omega z(x^1, x^2)}   Dz Dz^{\dagger}  Da \; e^{-S[z, \bar z, a]} \cr
&=  \int_{z(x^1, x^2)= z(x^1, x^2)}   Dz Dz^{\dagger}  Da \; e^{-S[z, \bar z, a, \Omega]} 
\end{align}
In the last step we used a field redefinition to replace $\Omega$-twisted boundary condition with  $\Omega$-holonomy. 
Several observations are in order:

\begin{itemize} 
\item Neither the  Hilbert space of the theory nor the degeneracies associated with states are altered in our graded construction.  Rather, we consider a graded state-sum with the insertion of  $\widehat{\Omega}$ into the partition function,  $\mathrm{tr}\left[\widehat{\Omega}\exp(-L \widehat{H})\right] $. We call this procedure {\it quantum distillation}. The distillation effectively picks out a subset of states, without applying any projections. 
The  graded state-sum has no simple thermal interpretation (at least associated with the original theory). 

\item In the presence of large global symmetries, quantum distillation can in principle over-whelm exponential Hagedorn growth in the Hilbert space.  In this sense, the graded partition function $\calZ_{\Omega}(L)$ has a much stronger change of being an analytic function of $L$, with the possibility of no phase transition as $L$ changes from small to large $L$. 

\item Due to asymptotic freedom,  the small-$L$ regime becomes weakly coupled and semi-classically accessible. What is learned on small-$L$ is not detached from $\R^2$. In circumstances where $\calZ_{\Omega}(L)$ is an analytic function of $L$,  the ground states of these two regimes are guaranteed to be continuously connected. This is the idea of adiabatic continuity. 

\item Anomaly matching implies that there are three possibilities in these theories in the IR, when the Coleman-Mermin-Wagner theorem is taken into account:  spontaneously broken discrete symmetry,  TQFT or CFT. This anomaly consideration in principle permits a phase transition between these possibilities. But provided  $\calZ_{\Omega}(L)$ is an analytic function of $L$, what takes place at weak coupling at small-$L$ also takes place at strong coupling at large-$L$. 

\item In the large-$N$  limit, we can prove volume independence for observables  provided they are measured using  $\calZ_{\Omega}(L)$.  
The fact that volume independence works implies that a distillation in the Hilbert space  may even overcome exponential growth in the density of states. 

\item
A similar graded Hilbert space analysis should apply to other 2d asymptotically free sigma models, such as the principal chiral model, and the $O(N)$ sigma model.
\end{itemize}

The overall construction suggests the appealing possibility  for analyticity of the graded partition function with important implications for non-perturbative properties of 2d QFT in the decompactification limit.

\acknowledgments
We are grateful to    Aleksey Cherman,  David Gross,  Tatsuhiro Misumi, Misha Shifman, 
Tin Sulejmanpasic, Sergei Gukov, Wolfgang Lerche, David Tong and  Arkady Vainshtein  for discussions.   
All three authors  thank  the KITP at Santa Barbara for its hospitality during the program 
``Resurgent Asymptotics in Physics and Mathematics'' where  this work was done. 
Research at KITP is supported by the National Science Foundation under
Grant No. NSF PHY-1125915. 
The work of Y.~T. is supported by the RIKEN Special Postdoctoral Researcher Program. 
This material is based upon work supported by the U.S. Department of Energy, Office of Science, Office of High Energy Physics under Award Number DE-SC0010339 (GD), and  by the U.S. Department of Energy, Office of Science, Office of Nuclear Physics under Award Number DE-FG02-03ER41260 (MU).

\bibliographystyle{utphys}
\bibliography{./QFT,./ref}
\end{document}